# Introduction to Chaos in Deterministic Systems

Carlos Gershenson
C.Gershenson@sussex.ac.uk

## 1. Introduction

The scope of this teaching package is to make a brief introduction to some notions and properties of chaotic systems. We first make a brief introduction to chaos in general and then we show some important properties of chaotic systems using the logistic map and its bifurcation diagram. We also show the universality found in "the route to chaos".

The user is only required to have notions of algebra, so it is quite accessible. The formal basis of chaos theory are not covered in this introduction, but are pointed out for the reader interested in them. Therefore, this package is also useful for people who are interested in going deep into the mathematical theories, because it is a simple introduction of the terminology, and because it points out which are the original sources of information (so there is no danger in falling in the trap of "Learn Chaos in 48 hours" or "Bifurcation Diagrams for Dummies").

The included exercises are suggested for consolidating the covered topics. The on-line resources are highly recommended for extending this brief induction.

## 2. Chaos

The study of chaotic dynamics in deterministic systems has become very popular in the past few decades, emerging from the study of **non-linear dynamics**. Perhaps it is because of the amazing findings that the study of chaotic systems has delivered. First, it would be natural to think that if a system is deterministic, its behaviour should be easily predicted. But there are systems where their behaviour turns out to be non-predictable: not because of lack of determinism, but because the complexity of the dynamics require a precision that is unable to be computed. This can be seen in systems where very similar initial conditions yield very different behaviours. Thus, let's say if we have the initial states 2.1234567890 and 2.1234567891, after some time the system will be for the first case in 3.5 and in the other in -1.7. So, no matter how much precision we have, the most minimal differences will tend in the long time to very different results. This is because there is an exponential divergence of the trajectories of the system (This can formally be measured with **Lyapunov exponents**).

Another interesting property of chaotic systems are **strange attractors**[1]. If one thinks of chaotic dynamics, it would be easy to assume that the dynamics follow no pattern. But if we look carefully there is an amazing pattern, in such a way that the states will not repeat themselves, but will be in a determined area of the state-space. And the properties of the strange attractors are astonishing, since they are **self-affine** (Mandelbrot, 1998) (a part of the attractor resembles the attractor), and therefore **fractals**.

---

[1]An attractor is a part of the state-space (which is the set of all possible states of a system) which "pulls" the dynamics into it. For example, a simple pendulum with friction has a **stable attractor** in the bottom of the vertical axis, because wherever the pendulum is, it will end at some time in that point. But there is an **unstable attractor** in the top of the vertical axis, because if that is the initial condition, in theory the pendulum would stay up there, but the most minimal perturbation would move the pendulum out of the unstable attractor. The unstable attractor repels the dynamics of the system.

## 3. Logistic map

One could think that chaotic systems need complicated formulae, but there are very simple functions which can lead not only chaos, but how this develops from "ordered" behaviour. The logistic function, used in population dynamics, is one of these functions, which we will describe in this section. The logistic function is

$$f(x) = ax(1-x)$$

where a indicates the "fertility" or "growth rate" of a population with limited resources. We can see that this function is a parabola, intersecting the x axis in (0,0) and (1,0).

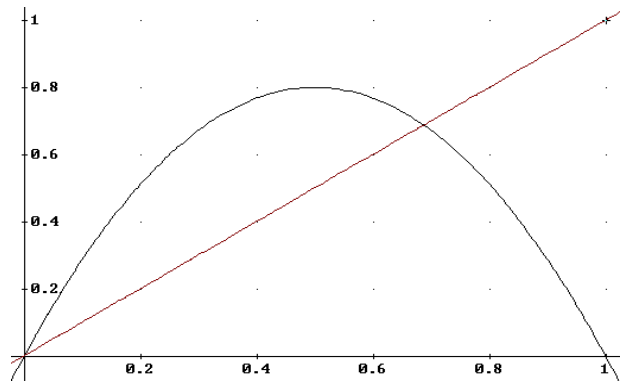
Figure 1. Logistic function (a=3.2)

For values 0<=a<=4, the height of the parabola will be in the interval [0..1]. If we **iterate** this function, we will observe the discrete dynamics of the population that the function models.

For iterating a function we need an initial value $x_0$, and the result of the function will be the next input of the function. Therefore

$$x_1 = f(x_0)$$

$$x_2 = f(x_1) = f^2(x_0)$$

…

$$x_n = f(x_{n-1}) = f^n(x_0)$$

where $x_n$ is called the $n^{th}$ iteration of $x_0$. The set of all the iterations of a function is called the **map** of the function.

An easy way to visualize the iteration of a function, is to plot the straight line y=x (also called identity line), and the function f(x). If we begin to draw lines following the points $(x_0,x_0)$, $(x_0,f(x_0)=x_1)$, $(x_1, x_1)$, $(x_1,f(x_1)=x_2)$, $(x_2, x_2)$, $(x_2,f(x_2)=x_3)$, … we can see that the iteration can be done geometrically just by drawing lines from f(x) to the line y=x and back to f(x).

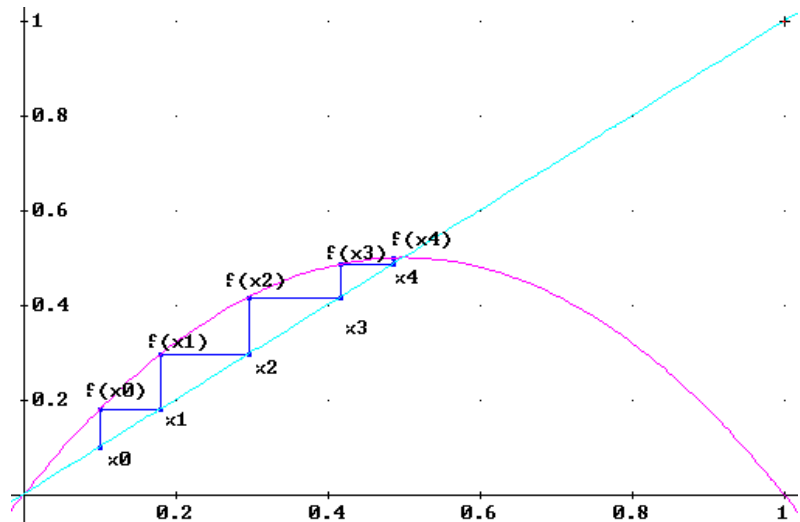
Figure 2. Iterating the logistic function (a=2.0).

For the logistic function, if $0<=a<=4$ and $0<=x_0<=1$, the dynamics will be confined to $[0..1]^2$.

Let's observe the dynamics in the logistic map as we increase the value of a.

For a=0, f(x)=0, independently of $x_0$.

If we increase a a bit, we can see that after some iterations, independently of $x_0$, the dynamics will reach $x_n=0$.

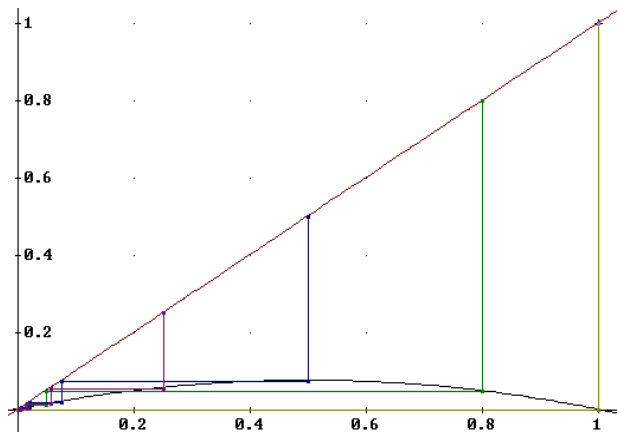
Figure 3. a=0.3, $x_0$=0.25, 0.5, 0.8, and 1.0.

For this case, 0 is a **stable attractor** of the map. If we increase a bit more a, we still observe this behaviour.

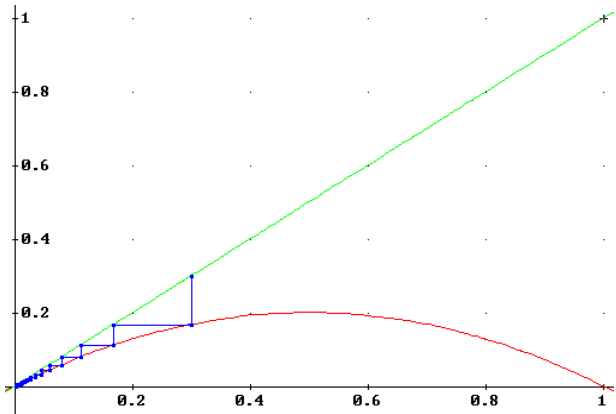
Figure 4. a=0.8, $x_0$=0.3.

Increasing a more, we can see that this behaviour changes somewhere.

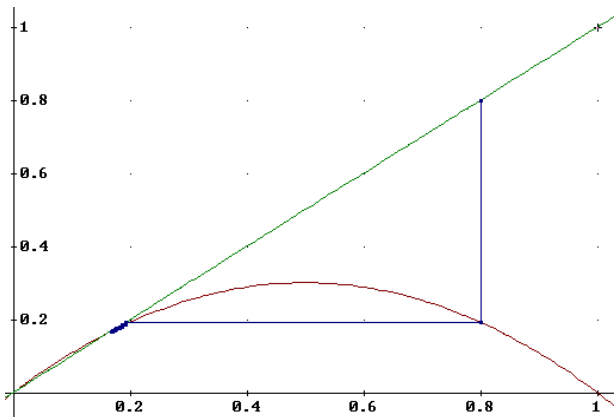
Figure 5. a=1.2, x0=0.8.

The attractor is zero no more, but the intersection of the parabola with the identity line. Where does this change occurs? We observe that it occurs when the parabola intersects the identity line in other point than the origin. And this transition is given when a=1.0.

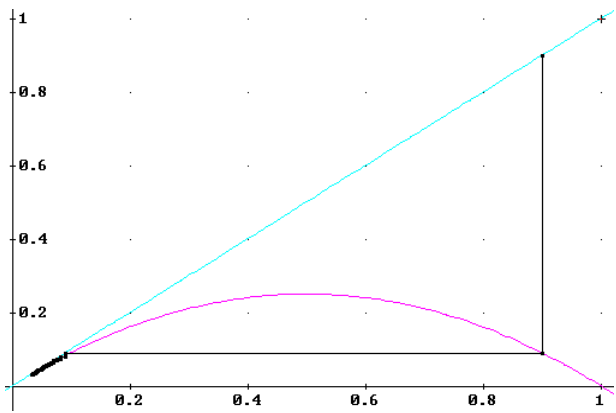

For a=1, the attractor 0 is asymptotic. This means that it is reached in the limit n→∞. But even for values of a>1, we can see that for values of $x_0$=0.0 or 1.0, 0 the dynamics will still be taken to zero. But if $x_0$ is very close to zero or one, the dynamics will be repelled. We can say that zero turned into an **unstable attractor**. Like an upright pendulum, if the initial condition is the unstable attractor, the system will stay in that state, but the most minimum perturbation will take the system away from that unstable state.

If we increase a, we still perceive the same behaviour: the dynamics will be attracted into the intersection of the parabola with the identity line.

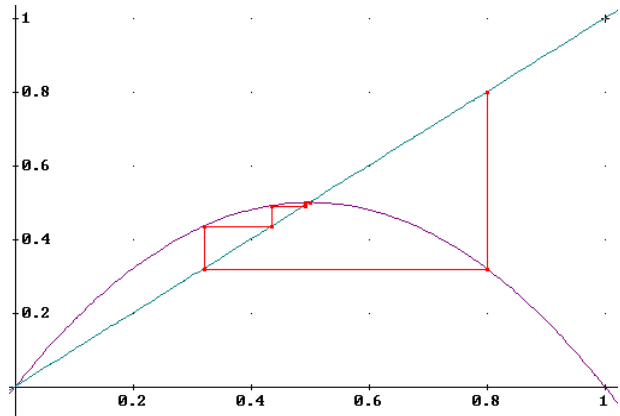

Figure 7. a=2.0, x0=0.8.

But the attractor point is changing. How can we calculate it? It is not hard. A point attractor satisfies $f(x_n)=x_n$. So we just need to find the roots of the equation

$$x = ax(1 - x)$$

which are x=0, x=1-1/a. We can see that the first one is stable for a<=1 and unstable for a>1. The second root is out of the interval [0..1] when a<1. When a=1, both roots are the same. But this second root is always stable? Let's plot some more maps.

If we increase a, we will see that the dynamics converge to the attractor point (1-1/a) more slowly as we approach a=3.0. And it is precisely at this point that we find something similar that at a=1.0.

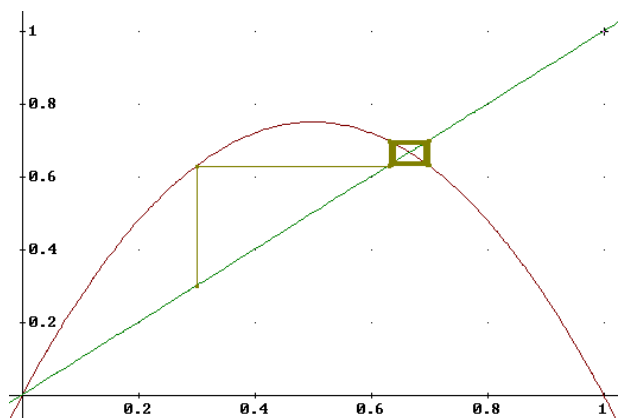

Figure 8. a-3.0, x0=0.3.

The dynamics tend asymptotically to the attractor, and will reach it on the limit $n\to\infty$. What happens if we increase a just a bit more?

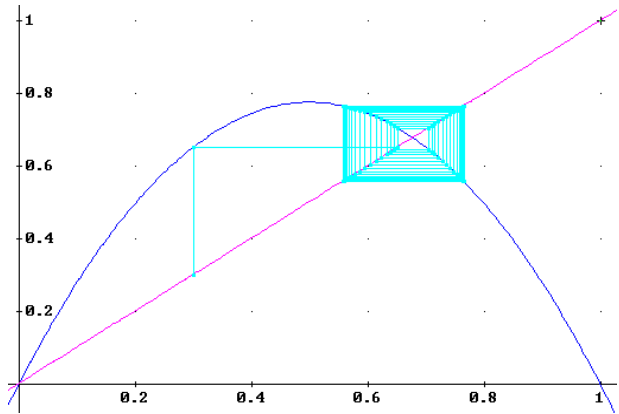
Figure 9. a=3.1, x0=0.3.

We can see that the point attractor became unstable, as it occurred with x=0. And now we have a **cycle attractor** (also called **orbit**) of period 2. This implies that $x_n = x_{n-2}$. This will be easier to visualize with $f^2(x)$:

$$f^{\,2}(x) = f(f(x)) = a(ax(1-x))(1-(ax(1-x)))$$

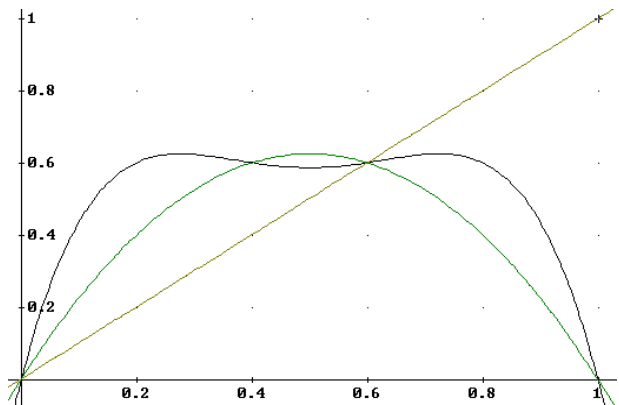
Figure 11. a=2.5.

For a<3, $f^2$ just intersects the identity line in 1-1/a, but precisely at a=3, we can observe a change.

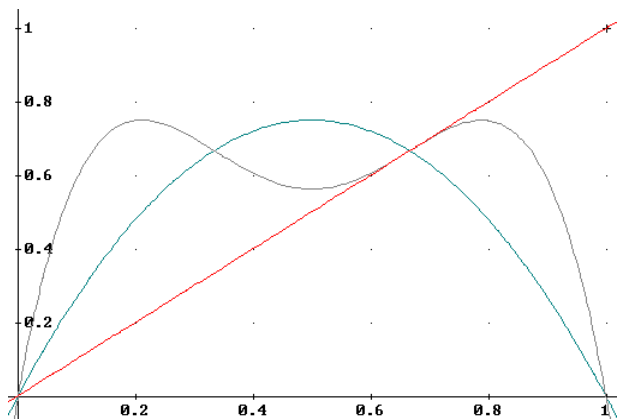
Figure 12. a=3.0.

By examining greater values of a, we can see that what occurs is that $f^2$ for a>3 intersects the identity line in three points.

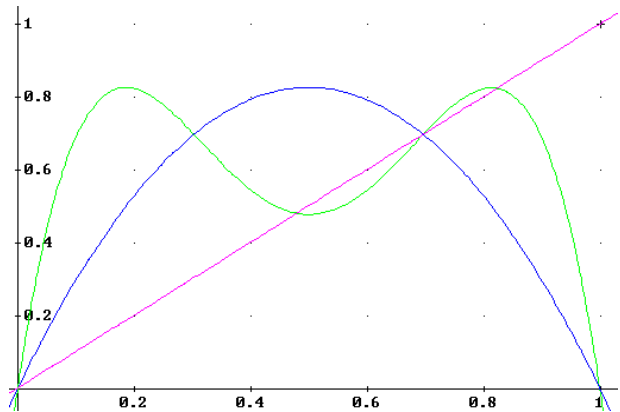
Figure 13. a=3.3.

And if we iterate the logistic function, we will see that the period 2 attractor is given precisely by the points where $f^2$ intersects the identity line.

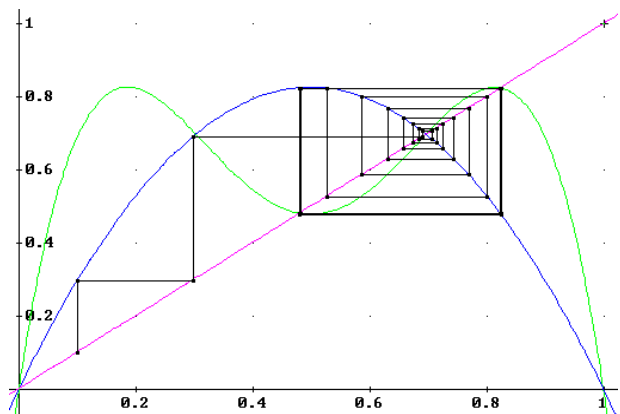
Figure 14. a=3.3, x0=0.1,0.696969...

And we can see that 1-1/a becomes an unstable attractor.

We can find again the points of the two-cycle attractor, solving the equation $f^2(x)=x$, where $f^2(x)=a(ax(1-x))(1-ax(1-x))$. It is a fourth degree equation, and the roots are x=0, x=1-1/a, x=$\sqrt{(a+1)}\sqrt{(a-3)}/2a + 1/2a + 1/2$, x=$-\sqrt{(a+1)}\sqrt{(a-3)}/2a + 1/2a + 1/2$. We can see that the first two are the same as for f(x), and the latter two are the other intersections. But do these become unstable as well?[2] Yes, actually, they do, and at the same time, and we will have an attractor of period 4.

---

[2] We can calculate if a point is stable or unstable if small perturbations grow or decrease with each iteration. This can be measured with the derivative $dx_{n+1}/dx_n = a(1-2x_n) = 2 - a$. If its absolute value exceeds one, then the point is unstable. If it is smaller than one, it is a stable point. If it equals one, it is a **bifuraction** point.

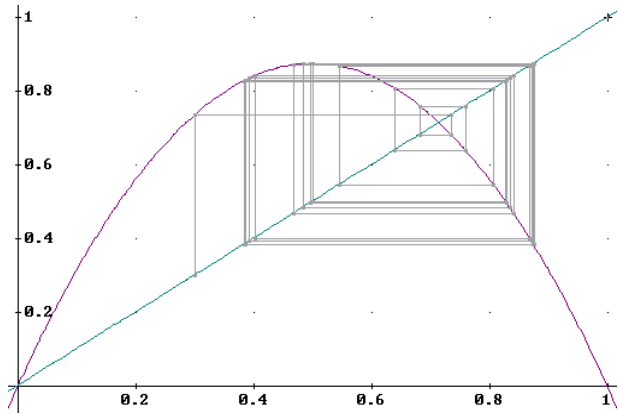
Figure 15. a=3.5, x0=0.3.

We can again find the points of the attractor of period 4 solving the equation $x=f^4(x)$, but this is beginning to get a bit complicated...

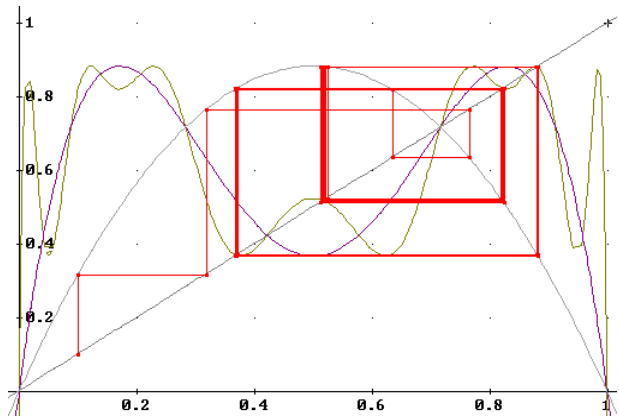
Figure 16. a=3.53, x0=0.1.

We can see that the attractor points are the four ones of $f^4$ that begin to intersect the identity line. The previous became unstable points. If we increase a just a bit more, we will find attractor cycles of period 8, 16, 32, 64, ... The values of a where there is a change of period (which is doubled), are called **bifurcation** points.

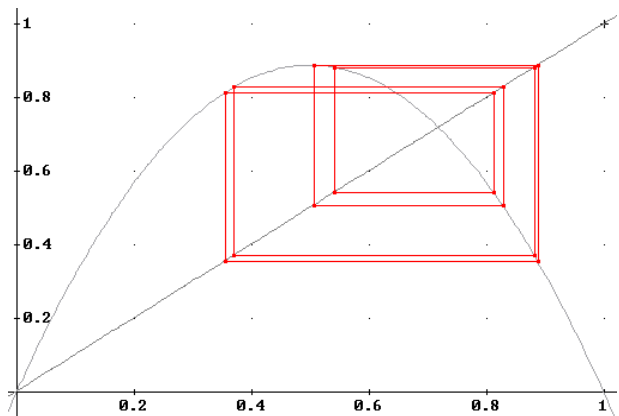
Figure 17. a=3.55, first iterations omitted.
Attractor cycle of period 8.

It gets extremely difficult to find these points analytically, so we will desist from trying to find the roots of $f^{1024}(n)$, and follow a different method. Moreover, if we increase a even more, we will find out that the dynamics seem not to be repeated. We have reached **chaos**. We can see it as an infinitude of stable and unstable points, each one pulling or pushing the dynamics making the behaviour... chaotic.

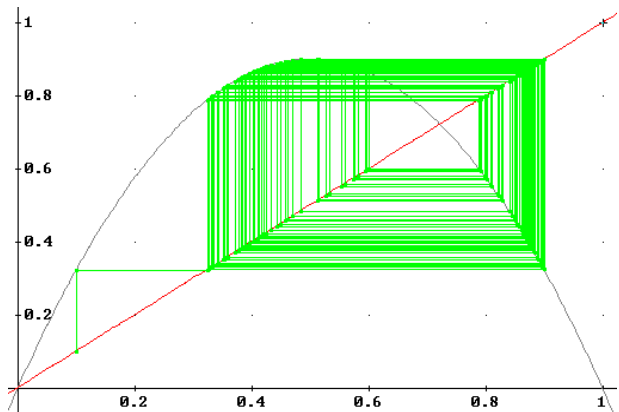
Figure 18. a=3.6, x0=0.1.

But we can see that the dynamics of the map are not random. They do not cover all the possible state-space (in this case, [0..1]). The area that the dynamics are concentrated in is known as a **strange attractor**.

## 4. Bifurcation Diagrams

Bifurcation diagrams are a very useful tool for observing what is going on. Basically, we will plot, the value of a against the points where the dynamic has concentrated after some initial iterations, this is, an attractor (point, cycle, or strange).

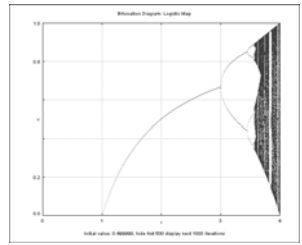

Here we can see with clarity all our previous observations and many more: For 0<a<1, 0 is a stable point attractor. a=1 is a bifurcation point, and 1<a<3 will have the attractor point 1-1/a. a=3 is another bifurcation point, and now there is a period 2 cycle attractor, that doubles to periods 4, 8, 16, 32, ... Let's take a closer look at this part of the bifurcation diagram.

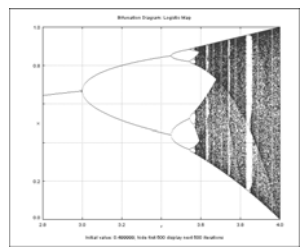

We can see that as the period of the attractors is doubled, the next bifurcation point is closer to the previous. After this limit n→∞, we could say that we have attractors of period "more than infinite". This is where the chaotic region begins. And when a=4, the strange attractor covers all the state space [0..1] (But note that there are areas with different densities...).

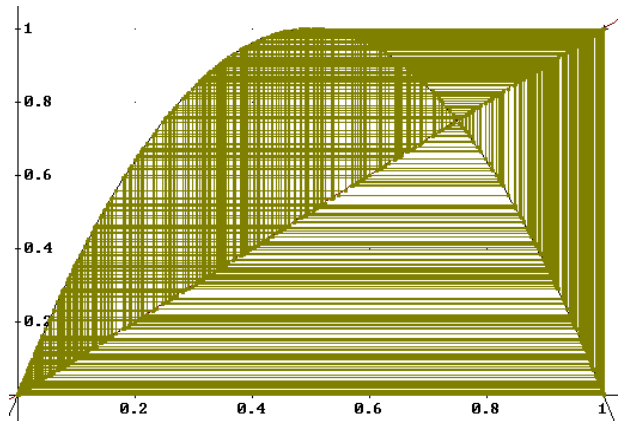
Figure 21. a=4.0, x0=0.1.

But in the bifurcation diagram we observe some regions where there are other periods, called **windows**. Notice a large window of period 3 around a=3.83. These periods also double to 6, 12, 24... and reach again a chaotic region.

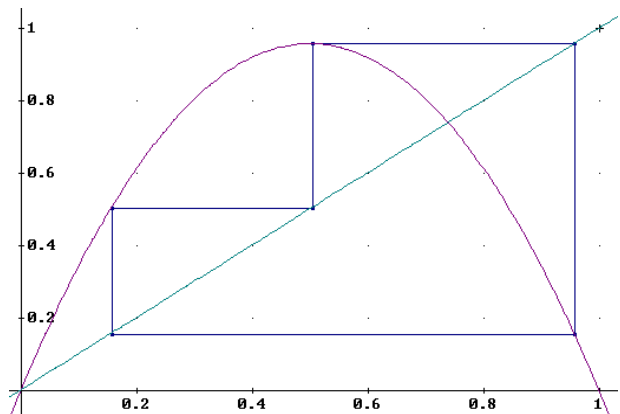
Figure 22. a=3.83.

But we can see that there are many other windows of different periods (emerging from **accumulation points**). Actually, it can be proven that in this bifurcation diagram are orbits of period n for all natural numbers n.

Let's take a closer look at the upper part of the first bifurcations.

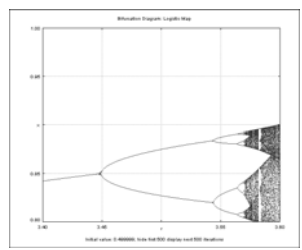

Let's zoom a bit more, following the upper bifurcations always.

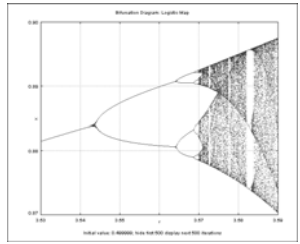

Hmmm... it is quite similar to the original bifurcation diagram. Perhaps a bit scaled. But we can see also a window of period three, and many other windows. And if we keep on zooming?

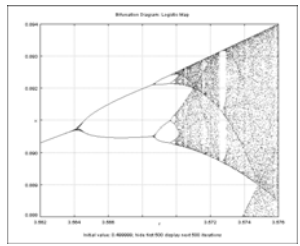

Just one more...

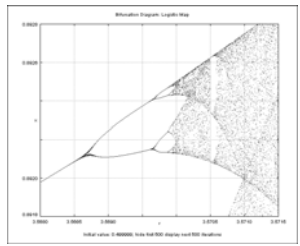

Well, this does not seem to get simpler. Actually, it looks like if the changes of scale would do nothing to the diagram. Yes, this is because this bifurcation diagram is a **fractal**. It is **self-affine** (Mandelbrot, 1998).

But what? We said that the first bifurcation diagram contained orbits of period of all the natural numbers. And if we zoom on a part of the bifurcation diagram, no matter how much we zoom, we will find orbits of all the natural numbers again! The natural numbers contain themselves infinitely number of times! Well, yes, they are infinite, but this is the conclusion is not obvious...

## 5. Universality

Around October, 1975, Mitch Feigenbaum was studying the properties of the logistic map with a programmable pocket calculator. As he saw that it took lots of strength to try to find the bifurcation points calculating every possible value of a (there were not so fast and not so many computers...), he studied the geometrical convergence of the doubling of periods, since it seemed to be some regularity.

So, we can call the first bifurcation point $a_0$, which for the logistic map is 1. The second bifurcation point $a_1=3$. And so, $a_2=3.4495...$, $a_3=3.5441...$, $a_4=3.5644...$ So, the ratio of change is

$$\frac{a_n - a_{n-1}}{a_{n+1} - a_n}$$

So, Feigenbaum (1983) calculated the limit

$$\lim_{n \to \infty} \frac{a_n - a_{n-1}}{a_{n+1} - a_n} = \delta = 4.6692016091029...$$

Basing himself on work on universal series by Metropolis, Stein, and Stein (1973), Feigenbaum found that the limit of ratios δ was the same for the sine map (a*sin(πx)), and actually, for any other function with doubling of periods. Thus, Feigenbaum's δ, turned out to be an universal constant. Independently of the function, δ will always be the ratio of the bifurcations leading to chaos. Feigenbaum's δ is also present in the famous Mandelbrot set

But Feigenbaum also found another universal constant. If you calculate the limit of the ratios of the distance from x=0.5 (which is the critical point in the logistic map), to the nearest point of the attractor cycle, it will also be an universal constant, called Feigenbaum's α. When a point of the attractor cycle equals 0.5, it is called a **superstable** orbit, because it is when it converges most quickly to the attractor.

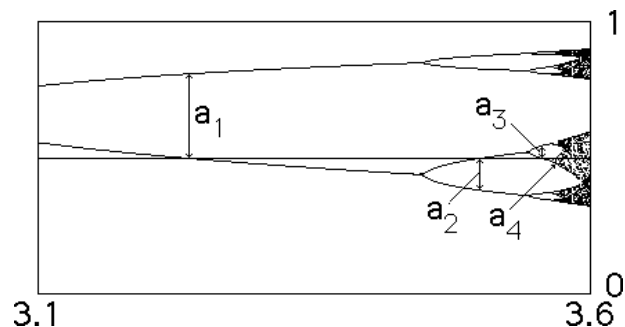

So we have

$$\lim_{n \to \infty} \frac{a_n}{a_{n+1}} = \alpha = 2.502907876...$$

The ratio of distances between the values of a where there are superstable orbits is also δ.

## 6. Exercises

This exercises are designed for the reader who knows how to use a mathematical assistant, such as Matlab, Mathematica, Derive, Mathcad, etc.

1. Plot the bifurcation diagram of the sine map f(x)=a*sin(πx). (0<a<1). What are the similarities with the bifurcation diagram of the logistic map?
2. Using the sine map, approximate Feigenbaum's constants α and δ. Tip: you need to calculate the bifurcation points first. Tip 2: you can actually use δ to approximate the bifurcation points and the values of a when a point of the attractor cycle=0.5 (superstable orbits).

## 7. Further reading

- Lorenz, Edward N. (1993). *The Essence of Chaos*, Univ. Washington Press. Seattle.
- Mandelbrot, Benoit. (1998). *Multifractals and 1/f Noise: Wild Self-affinity in Physics (1963-1976)*. Springer. New York.
- Hofstadter, Douglas R. (1985). Metamagical Themas: Questing for the Essence of Mind and Pattern. Basic Books. New York.
- Elert, G. (1995). The Chaos Hypertextbook. Http://hypertextbook.com/chaos

## 8. Online resources

- There is the sci.nonlinear newsgroup, and their FAQ can be found at http://amath.colorado.edu/faculty/jdm/faq.html
- Books:
    - Elert, G. (1995). The Chaos Hypertextbook. Http://hypertextbook.com/chaos
    - Strange Attractors: Creating Patterns in Chaos, by Julien C. Sprott. http://sprott.physics.wisc.edu/sa.htm
- Software for studying chaos:
    - Exploring Chaos: Theory and Experiment, by Brian Davies. The bifurcation diagrams presented here were generated using a small part of his free software. http://wwwmaths.anu.edu.au/~briand/chaos/
    - The Period Doubling Route to Chaos, by Michael Cross, with great Java Applet. http://www.cmp.caltech.edu/~mcc/chaos_new/Scalemap.html
- Applied Chaos Tutorial, http://www.physics.gatech.edu/chaos/tutorial/outline.htm

## 10. Appendixes

### 10.1. Answers to Exercises

These solutions are for Derive, but the syntax is explicit enough to allow their translation to other mathematical assistants.

1. The following functions are used to plot the bifurcation diagram of the sine map. The properties of the bifurcation diagram are the same than the ones found in the one of the logistic map: the attractor cycles of period n occur in the same order.

"The sine function"
S(a, x) := a·SIN(πx)
"Vector of iterations of S(a,x)"
MAP(a, x0, n) := ITERATES(S(a, x), x, x0, n)
"Auxiliary for plotting the points (a,iteration)"
PLOTMAP(a, v) := VECTOR([a, ELEMENT(v, i)], i, 1, DIMENSION(v))
"Just get the elements from m to n (discard first m)"
PERIODICMAP(a, x0, m, n) := VECTOR(ELEMENT(MAP(a, x0, n), i), i, m, n)
"Calculate trajectories each da from a=0 to 1, with initial state x0, taking first m iterations, and keeping following n–m iterations"
BIFURATION_DIAGRAM(da, x0, m, n) := VECTOR(PLOTMAP(i, PERIODICMAP(i, x0, m, n)), i, 0, 1, da)
"Examples (as da gets smaller, it takes more time...but the plot is nicer...)"
BIFURATION_DIAGRAM(0.1, 0.4, 50, 60)
BIFURATION_DIAGRAM(0.01, 0.4, 100, 120)
BIFURATION_DIAGRAM(0.002, 0.4, 100, 120)

2. We find by approximation the values of a for superstable orbits of periods 2, 4, 8, 16...

"Iteration n of the sine function"
SN(a, x0, n) := ITERATE(S(a, x), x, x0, n)
"Recursive function for trying to find a values for superstable orbits (when an element of the cycle equals 05) of period n, in steps of δ, starting with initial a"
SUPEREST(a, δ, n) := IF(SN(a, 0.5, n) = 1/2, a, SUPEREST(a + δ, δ, n)|
"Let's find the values (cheat using Feigenbaum's constant to get a nice initial a)"
SUPEREST(0.48, 0.000001, 1)
0.5
SUPEREST(0.77, 0.000001, 2)
0.777727
SUPEREST(0.84, 0.000001, 4)
0.846379
SUPEREST(0.86, 0.000001, 8)
0.861448
SUPEREST(0.864, 0.000001, 16)
0.864692
SUPEREST(0.865, 0.000001, 32)
0.865389
"Approximation to Feigenbaum's δ"
 0.864692 - 0.861448 / 0.865389 - 0.864692
4.65419
"Let's put our a's for superstable orbits in a vector..."
d := [0.5, 0.777727, 0.846379, 0.861448, 0.864692, 0.865389]
"Function to get attractor cycles of length n (supposing x0 is part of the attractor cycle)"
SNI(a, x0, n) := ITERATES(S(a, x), x, x0, n)
"Let's see..."
SNI(0.777727, 0.5, 2)
[0.5, 0.777726, 0.500008]
SNI(0.846379, 0.5, 4)
[0.5, 0.846379, 0.392802, 0.798834, 0.5]
SNI(0.861448, 0.5, 8)
[0.5, 0.861447, 0.363236, 0.78315, 0.54251, 0.853777, 0.381953, 0.802885, 0.500007]
SNI(0.864692, 0.5, 16)

[0.5, 0.864691, 0.356595, 0.778414, 0.554487, 0.852054, 0.38758, 0.811322, 0.483053, 0.863466, 0.359625, 0.781962, 0.547057, 0.85526, 0.379777, 0.803747, 0.499981]
SNI(0.865389, 0.5, 32)
[0.5, 0.865389, 0.355155, 0.777329, 0.557194, 0.851456, 0.389345, 0.813623, 0.478242, 0.863368, 0.360159, 0.783211, 0.544864, 0.856807, 0.3763, 0.800861, 0.506767, 0.865193, 0.355641, 0.777908, 0.555989, 0.852036, 0.387937, 0.812311, 0.48121, 0.863881, 0.358888, 0.781735, 0.547975, 0.855578, 0.379306, 0.803922, 0.5]
"Let's calculate the distances from 0.5 to the closest point..."
as := [0, 0.777727 - 0.5, 0.5 - 0.392802, 0.54251 - 0.5, 0.5 - 0.483053, 0.506767 - 0.5]
[0, 0.277726, 0.107198, 0.0425099, 0.0169469, 0.006767]
"Approximation to Feigenbaum's α"
0.0425099 - 0.0169469 / 0.0169469 - 0.006767
2.51112

## 10.2. Code

Here we present the code of Derive used to generate the figures of the logistic map.

"Generate the logistic function for the specified a, and the identity line, for plotting"
LOGMAP(a) := [y = a·x·(1 - x), y = x]
"Generate the points from the vector v..."
MAPVEC(v) := APPEND(VECTOR([[ELEMENT(v, i), ELEMENT(v, i)],[ELEMENT(v, i), ELEMENT(v, i+1)]], i, 1, DIMENSION(v) - 1))
"...to get points from the identity to the parabola."
LOGVEC(x0, a, n) := MAPVEC(ITERATES(a·x·(1 - x), x, x0, n))
"examples"
LOGVEC(0.3, 2.4, 20)
LOGMAP(2.4)
"Generate $f^2$ for plotting"
LOGMAP2(a) := [y = a·(a·x·(1 - x))·(1 - a·x·(1 - x)), y = x]
"Generate $f^4$ for plotting"
LOGMAP4(a) := [y = a·(a·a·(a·x·(1 - x))·(1 - a·x·(1 - x))·(1 - a·(a·x·(1 - x))·(1 - a·x·(1 - x))))·(1 - a·a·(a·x·(1 - x))·(1 - a·x·(1 - x))·(1 - a·(a·x·(1 - x))·(1 - a·x·(1 - x)))), y = x]